\documentclass[12pt]{iopart}
\usepackage{graphicx}
\newcommand{\la}{\lambda}
\newcommand{\om}{\omega}
\newcommand{\ep}{\epsilon}
\newcommand{\de}{\delta}
\newcommand{\bea}{\begin{eqnarray}}
\newcommand{\beq}{\begin{equation}}
\newcommand{\eea}{\end{eqnarray}}
\newcommand{\eeq}{\end{equation}}
\begin{document}
\title{The physics of exceptional points}

\author{W.D.~Heiss}

\address{National Institute for Theoretical Physics, Stellenbosch
  Institute for Advanced Study}

\ead{dieter@physics.sun.ac.za}

\begin{abstract}
A short resume is given about the nature of exceptional points (EPs) followed by
discussions  about their ubiquitous occurrence
in a great variety of physical problems. EPs feature in classical as well as in quantum
mechanical problems. They are associated with symmetry breaking for 
${\cal PT}$-symmetric Hamiltonians, where a great number of experiments have been performed
in particular in optics, and to an increasing extent in atomic and molecular physics.
EPs are involved in quantum phase transition and quantum chaos,
they produce dramatic effects in multichannel scattering, specific time dependence and more.
In nuclear physics they are associated with instabilities and continuum problems.
Being spectral singularities they also affect approximation schemes.
\end{abstract}
\maketitle

\section{Introduction}
Singularities of functions describing analytically observable quantities
have always been at the scrutiny of theoretical investigations. For instance,
the structures of measured cross sections are usually associated with
pole terms in the complex energy plane of the scattering amplitudes.
In turn, these pole terms are associated with specific boundary conditions
of solutions of, say, the Schr\"odinger equation \cite{newt}.
Another example is
the pattern of spectra when plotted versus an external strength parameter, say,
of a magnetic field; it usually shows the phenomenon of level repulsion, often
associated with quantum chaos \cite{wintg}. When such spectra are continued
into the complex plane of the strength parameter, one encounters
a different type of singularities where two repelling levels
are connected by a square root branch point. If for real strength parameter
the Hamiltonian is hermitian, the branch points always occur at complex
parameter values thus rendering the continued Hamiltonian as non-hermitian.
As a consequence, the well-known properties associated with a degeneracy
of hermitian operators are no longer valid. These singularities have been
dubbed {\em exceptional points} (EPs) by Kato \cite{kato}. 

A few decades ago these singularities have been perceived as a mathematical
phenomenon that would be 'in the way' of approximation schemes \cite{he0,mois}
(see also the discussion in Section 4 below). However, their physical significance
has been recognised in an early paper by Berry \cite{ber2} based on an observation
by Pancheratnam \cite{panch}. The specific algebraic property of the dielectric
tensor (it cannot be diagonalised) brings about particular physical effects.
This has been expounded in great detail in \cite{ber3} for particular optical systems.
As is discussed below optical systems constitute one major realm where EPs have a
major bearing. Yet, originating from a parameter dependent eigenvalue problem
EPs naturally occur and can give rise to dramatic effects in a great variety of
physical problems: in mechanics, electro-magnetism, atomic and molecular physics,
quantum phase transitions, quantum chaos and more.

In Section 3 we return to these phenomena in detail. In 3.1 we discuss the experimental
manifestation using microwave cavities where all the mathematical properties have been
confirmed to be a physical reality. The investigation and physical realisation
of $\cal PT$-symmetric Hamiltonians has become a major subject of theoretical and
experimental endeavor. An important role is played by EPs in this research, it is dealt with in 3.2.
Subsection 3.3 is devoted to a short discussion of the role of EPs in atomic and molecular
physics while in 3.4 recent findings in open systems and laser physics are mentioned. 
In 3.5 we demonstrate the essential role of EPs in understanding the dramatic
effects associated with a quantum phase transition and quantum chaos, while 3.6 presents
a short discussion of multichannel scattering and 3.7 touches upon EPs of higher order.
Classical mechanics and fluids are briefly dealt with in 3.8.
The sometimes adverse role of EPs in approximation schemes is, for historical reasons,
discussed in Section 4.

For the sake of completeness we rehash the essential formal background in the following Section 2.
The last Section 5 gives a summary.

\begin{figure}
\includegraphics[height=0.18\textheight]{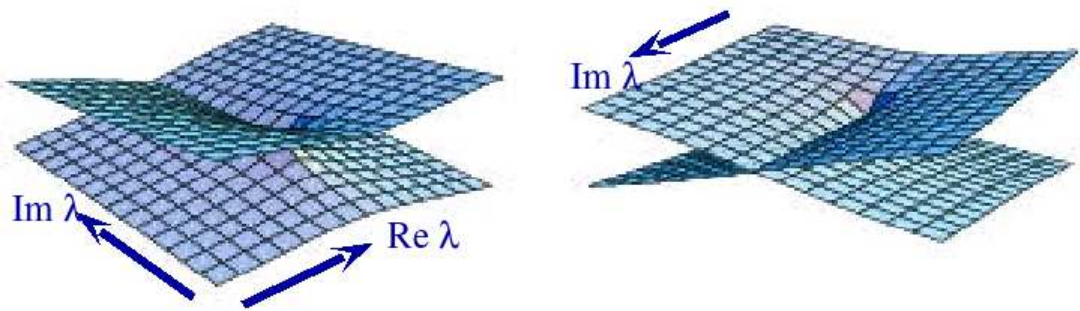}
\caption{Perspective view of the Riemann sheet structure of two
coalescing energy levels in the complex $\la-$plane. The assignment for
the axes assumes all other parameters as real; if some or all of them are complex
the sheet structure remains but is shifted and/or turned.}
\end{figure}

\section{Exceptional points}
Exceptional points occur generically in eigenvalue problems that depend
on a parameter. By variation of such parameter (usually into the complex plane)
one can generically find points where eigenvalues coincide. 

In the immediate vicinity of an EP the special algebraic behaviour
-- as discussed below -- allows a
reduction of the full problem to the two dimensional problem associated with
the two coinciding levels \cite{he}. We thus confine our discussion to
the eigenvalues of a two-dimensional matrix where the direct connection of an EP
and the phenomenon of level repulsion
is easily demonstrated. Consider the problem
\bea
H(\la )&=& H_0 +\la V \nonumber\\
&=&
\pmatrix{\om_1 & 0 \cr 0 & \om_2 }
+\la \pmatrix{\ep_1 & \de_1 \cr \de_2 & \ep_2}
\label{ham}
\eea
where the parameters $\omega_k$ and $\epsilon_k$ determine the
non-in\-ter\-ac\-ting resonance
energies $E_k=\om_k+\la \ep_k, \,k=1,2$. 
We may choose all parameters complex and we require $[H_0,V]\ne 0$ 
to ensure the problem to be non-trivial. Owing to the
interaction invoked by the matrix elements $\de_k$ the two levels
do not cross but repel each other. However, the two levels {\it coalesce}
at specific values of $\la $ in the vicinity of the level repulsion, that is at the two EPs
\bea
\la_1&=&\frac {-i(\om _1-\om _2)}{ i(\ep _1-\ep _2)+2 \sqrt{\de_1 \de_2} }\\ 
\la_2&=&\frac {-i(\om _1-\om _2)}{ i(\ep _1-\ep _2)-2 \sqrt{\de_1 \de_2}}.
\eea
For $\de_k\ne 0$ the energy levels have a square root singularity as a function of $\la $
and read
\beq
E_{1,2}(\la)=\frac{1}{2}\bigg(\om_1+\om_2+\la (\ep_1+\ep_2)
\pm \sqrt{(\ep_1-\ep_2)^2+4\de_1\de_2}\sqrt{(\la-\la_1)(\la-\la_2)}\bigg)
\eeq
and the eigenvalues at the EPs are
\beq
E(\la_{1,2})=\frac{\ep_1 \om_2-\ep_2 \om_1\mp i\sqrt{\de_1 \de_2}(\om_1+\om_2)}
{ \ep _1-\ep _2\mp 2 i \sqrt{\de_1 \de_2}}.
\eeq

We use the term {\it coalesce} as the pattern is distinctly different
from a degeneracy usually (but not only) encountered for hermitian operators. Note that
$H(\la )$ is hermitian only when $\om_k$, $\ep_k$ and $\la $ are real and
$\de_1 = \de_2^*$. At the EP the difference between a degeneracy and a coalescence
is clearly manifested by the occurrence of only {\em one} eigenvector
instead of the familiar two in the case of a genuine degeneracy. Using the
bi-orthogonal system for a non-hermitian matrix,
the (only one) right hand eigenvector reads at $\la =\la_1$ (up to a factor)
\beq
|\phi _1\rangle = \pmatrix {\frac{+i \de_1}{\sqrt{\de_1 \de_2}} \cr 1} 
\eeq
and at $\la=\la_2$
\beq
|\phi _2\rangle = \pmatrix {\frac{-i \de_1}{\sqrt{\de_1 \de_2}} \cr 1} ,
\eeq
while the corresponding left hand eigenvector is at $\la_1$ and $\la_2$
\bea
\langle \tilde \phi _1| &=& ( \frac{+i \de_2}{\sqrt{\de_1 \de_2}} , 1) \\
\langle \tilde \phi _2| &=& ( \frac{-i \de_2}{\sqrt{\de_1 \de_2}} , 1) ,
\eea
respectively.
Note that the norm -- that is the scalar
product $\langle \tilde \phi _k|\phi _k\rangle ,\; k=1,2$ -- vanishes which is often
referred to as self-orthogonality \cite{moisb}. It is this vanishing of the norm that enables
the reduction of a high dimensional problem to two dimensions in close
vicinity of an EP \cite{heha}.
The existence of only one eigenvector with vanishing norm is related to the fact that
for $\la=\la_1$ or $\la=\la_2$ the matrix $H(\la )$ cannot be diagonalised \cite{kato}.
At these points the Jordan decomposition reads
\beq
H(\la_1)=S \pmatrix{E(\la_1) & 1 \cr 0 & E(\la_1)}S^{-1}
\eeq
with
\beq
S=\pmatrix{\frac{i\de_1}{\sqrt{\de_1 \de_2}} & 
\frac{2i\sqrt{\de_1 \de_2}-\ep_1+\ep_2}{( \om_1- \om_2)\de_2}  \cr
1 & 0}.
\eeq
Similar expressions hold at $\la =\la_2$. We mention that the second column of $S$ is often referred to
as an associate vector obeying $(H(\la_1)-E(\la_1))|\psi_{assoc}\rangle = |\phi_1\rangle $.

If one or both $\de_k$ vanish there is level crossing ($\la _1=\la _2$). If only $\de_1$ or $\de_2$ vanishes 
then $V$ and thus $H(\la )$ cannot be diagonalised at the crossing point,
therefore the Jacobian form is non-diagonal and only
one eigenvector exists at the crossing point; yet it is not an EP as there is no
square root singularity in $\la $. We do not further pursue such case as it appears
to be of no physical interest.

The square root singularity also affects the the Green's function and thus the scattering matrix
in a particular way.
At the EPs the Green's function (and scattering matrix) have a pole of second order \cite{mondr}
in addition to the familiar pole of first order.
For the case considered here the explicit form of the Green's function reads
at $\la_1$
\beq
(E-H(\la_1))^{-1}=\frac {1}{E-E(\la_1)}\pmatrix{1 & 0 \cr 0 & 1}+
\frac {i\sqrt{\de_1\de_2}\la_1}{(E-E(\la_1))^2}\pmatrix{1 & i\sqrt{\frac{\de_1}{\de_2}} \cr
i\sqrt{\frac{\de_2}{\de_1}} & -1}
\eeq
and similar at $\la_2$. The first term resembles the conventional expression at a non-singular point of
the spectrum; the second term is a consequence of the singular spectral point. This term can give rise to
dramatic effects near to the EP \cite{HN10}, and at the EP special effects are generated from the
interference of the two terms when the square of the modulus of the
scattering matrix is considered.

The square root behaviour as given by Eq.(4) has further physical consequences as discussed in the following
sections. Here we list the major mathematical reasons giving rise to special physical phenomena.
\begin{itemize}
\item{} In the vicinity of an EP the spectrum is strongly dependent on the interaction parameter;
in fact, the derivative with respect to  $\la $ of the eigenvalues and eigenvectors is infinity at the EP.
\item{} Encircling $\la_1$ or $\la_2$ in the $\la -$plane will interchange the two energy levels (see Fig.1).
\item{} Avoiding the EP there are two linearly independent eigenvectors 
$|\psi_1\rangle $ and $|\psi_2\rangle $ (and the left hand companions 
$\langle \tilde \psi_1 | $ and $\langle \tilde \psi_2 | $)
with the normalisation $\langle \tilde \psi _k|\psi _k\rangle =1$.
Enforcing the normalisation into the EP
(where $\langle \tilde \psi _k|\psi _k\rangle $  vanishes) the components of the eigenvectors tend to
infinity as $\sim 1/(\la-\la_k)^{1/4}$. Repeated encircling -- say counterclockwise --
of an EP generates the pattern for the normalised eigenvectors
\beq
|\psi_1\rangle \to -|\psi_2\rangle \to -|\psi_1\rangle  \to |\psi_2\rangle \to |\psi_1\rangle 
\eeq
from which the pattern for $|\psi_2\rangle $ follows accordingly  \cite{he,ali};
the fourth root behaviour is clearly seen
as four rounds are needed to reach the initial sheet.
Notice that these relations imply a kind of chiral behaviour: going clockwise instead of
counterclockwise gives a different sign.
We note that these relations are presented inconsistently in \cite{rott}.
\item{} When the eigenfunctions $|\psi_1\rangle $ and $ |\psi_2\rangle $  coalesce
as given in (6) and (7) (here we normalise the vectors by setting one component equal to unity), 
they become, for $\de_1=\de_2$, independent
of parameters and assume the form $$ \pmatrix{\pm i \cr 1}. $$ The phase difference 
of $\pi/2$ between the two components is changed
if, for instance, time reversal symmetry is broken by choosing complex $\de_1=\de_2^*$ \cite{darm3,hahe}. 
Note further that $\la_k$ must be complex for hermitian operators $H_0$ and $V$  
in which case the two EPs occur at complex conjugate
values, i.e.~$\la_1=\la_2^*$. In general $\la_1$ or $\la_2$ can be real \cite{heis}.
Recall that an EP can be approached in
the laboratory only if the corresponding energy has a non-positive imaginary part.
\end{itemize}

\section{Physical effects}
Many cases of particular effects have been reported in the literature
during the past ten years. We here discuss only some major trends and
developments. While we focus the discussion upon quantum mechanical problems
and optics, the ubiquitous character of EPs in any parameter dependent eigenvalue
problem makes them appear also in problems of classical mechanics and others.

\subsection{Microwave cavity}
Probably for the first time ever the direct encircling of the square root
branch point - that is the manifestation of the two Riemann sheets
(see Fig.1) - was accomplished with a microwave resonator \cite{darm1}.

The realisation of the complex parameter $\la $ was implemented in the
laboratory by two real parameters: (i) the coupling between the two
halves of the cavity and (ii) the variation of the one level in one
half of the cavity. In one experiment the direct approach of the EP
was avoided while the encircling was done at close distance.
All properties as listed in the previous section have been confirmed:
encircling an EP swaps the energies and fourfold encirclements 
yield relations (13).

The chiral property of the wave function at the EP
has been confirmed in a second experiment \cite{darm2} where the phase
difference of $\pi/2$ between the two components has 
been measured in a direct approach of an EP.

The same results have been established with two coupled electronic
circuits \cite{timo}.

More recent experiments implementing carefully time reversal symmetry breaking
by using a magnetised device at the coupling of two cavities \cite{darm3} are
in perfect accordance with the predictions made in \cite{hahe}.

Other types of experiments and/or theoretical investigations with
microwave cavities are found in \cite{lee,wiers} where the effects of EPs
feature prominently. The study of continuous variation of the coupling between a cavity
and a rubidium atom enabled the direct observation of an EP in an open quantum
system \cite{ychoi}.

\subsection{${\cal PT}$-symmetric  Hamiltonians}
It has been suggested to extent the class of the traditional hermitian
Hamiltonians by a specific choice of non-hermitian
operators \cite{cmb1}. Hamiltonians that are symmetric under the combined
operation of parity and time reversal transformation (${\cal PT}$-symmetric
operators) can have a real spectrum even though the operators can be
non-hermitian. If the eigenstates are also
symmetric under ${\cal PT}$, that is if 
${\cal PT}|\psi _E\rangle = {\rm const}|\psi _E\rangle$,
the eigenvalues are real; when the symmetry is
broken, that is if the above relation does not hold, the
eigenvalues are complex \cite{dor}. It turns out that the parameter values
where the symmetry breaking occurs are just those values where an EP of the
system appears. Since the Hamiltonians considered in this class can be
non-hermitian, the EPs can occur for real parameter values. Obviously,
this particular extension of traditional quantum mechanics has attracted great
interest in the literature with umpteen theoretical and experimental papers.

Before turning to a selection of these we rehash some basic features of the specific
class of operators considered here. 
While unbounded non-Hermitian operators constitute a
difficult mathematical problem in general, we here deal with the much simpler
class that is well understood: the {\em quasi-Hermitian} operators \cite{sgh}.
A non-Hermitian operator $H\ne H^{\dagger}$ is called quasi-Hermitian
if there exists a bounded Hermitian positive-definite operator 
$\Theta $ that ensures the relation
\beq
\Theta H=H^{\dagger }\Theta.  
\eeq
The relation implies that $H$ is similarly equivalent to a Hermitian
operator, in fact the operator
\beq
h_S=SHS^{-1} \label{real}
\eeq
is Hermitian when $S$ is the positive root of $\Theta $. The
operator $\Theta $ may be viewed as a metric characterising a
different Hilbert space by defining the scalar product 
\beq
\langle \cdot |\cdot \rangle _{\Theta }:=\langle \cdot|\Theta \cdot
\rangle \label{scal}
\eeq
where $\langle \cdot|\cdot \rangle $ is the usual scalar
product, employing the $L^2$-metric being the identity. Two important
observations can be made: (i) the non-hermitian operator $H$ is in fact
hermitian if the new metric is used for the underlying Hilbert space; note,
however, that then other operators (like position or momentum) may no longer be
necessarily hermitian. There is freedom though in the choice of the
metric $\Theta $ \cite{sgh,ghs}; (ii) in view of (\ref{real}) the spectrum
of $H$ is real.

Of particular interest in our context is the fact that at a parameter value
where the symmetry gets broken -- at the EP --, the singularity affects the
the metric as well; in fact the metric ceases to exist \cite{musum}. While
the occurrence of an EP has been shown to cause a great variety of physical
effects, the break down of the metric has been associated with a speculative
suggestion: that the transition time between the two states connected at the EP
may be much shorter than the expected time $\sim \hbar /\Delta E$ \cite{brach}.

Many theoretical papers have been published on the subject dealing with the
mathematical aspects of diagonalisable non-hermitian
operators \cite{mostaf} - \cite{gu}; the list given here
can only be incomplete while the quoted papers contain further pertinent references.
More recent theoretical papers deal with specific subjects such as ${\cal PT}$-symmetry
in optics \cite{jones} and non-linear wave equations \cite{fring}.
In \cite{pros} a new class of chaotic systems with
dynamical localization is studied: a gain/loss parameter invokes a spontaneous phase 
transition from real values of the spectrum (phase of conserved symmetry) to complex values 
(phase of broken symmetry).

Perhaps some of the most beautiful demonstrations of ${\cal PT}$-symmetry breaking
at an EP have been made in optical systems. Light propagation is used in distributed-feedback 
optical structures with gain or loss regions. The
EPs at the points of ${\cal PT}$-symmetry breaking of the Dirac Hamiltonian 
give rise to simple observable physical quantities 
such as resonance narrowing and laser oscillation \cite{long}.
In ${\cal PT}$-symmetric optical lattices the transition at the spectral singularity (EP)
has been demonstrated \cite{makr} (see also \cite{guo}-\cite{bendix}). More experimental work
along these lines can be expected. In particular specific quantum mechanical systems
like the ones discussed in the following subsection are expected to involve aspects of
${\cal PT}$-symmetry.

An interesting twist is the classical analogue to quantum mechanical ${\cal PT}$-symmetric
Hamiltonians \cite{cmbclass}. It appears that the correspondence principle can be usefully
extended into the complex domain of classical variables. The association between complex 
quantum mechanics and complex classical mechanics is subtle and requires a non-perturbative
approach (see also \cite{musum,alimost}).

Quite generally and in the spirit of ${\cal PT}$-symmetry, its consideration is of course not
restricted to quantum systems. Here we mention electronic circuits where the connection of
instabilities in ${\cal PT}$-symmetric systems and EPs is investigated experimentally
\cite{schind} and theoretically \cite{kirill}.

\subsection{EPs in atomic/molecular physics, Feshbach resonances}
Using Feshbach resonance techniques
there are recent proposals for resonant dissociation by lasers of
$H_2^+$ molecules or alkali dimers, where the effects of EPs are
expected to feature prominently \cite{lef,moisa}. Similar in spirit,
a Bose-Einstein condensate of neutral atoms with induced electromagnetic 
attractive ($1/r$) interaction has been discussed recently as another system 
allowing a tunable interaction \cite{wun2}.  The critical value - an
EP - where the onset of
the collapse of the condensate occurs is interpreted as    
a transition point  from separate atoms to the formation 
of molecules or clusters \cite{wun}. In this context we recall that
level splitting by potential barriers (quantum tunneling) is associated
with a coalescence of the two levels at an appropriate (complex) value of
the barrier strength.

In a recent investigation 
it was found that by transporting an eigenstate around an EP
the final eigenstate may be different from the initial eigenstate. 
The precise interplay between gain/loss and non-adiabatic couplings
imposes specific limitations on the observability of this flip or non-flip
effect \cite{uzdin}.

\subsection{EPs in laser physics and open systems}
EPs can strongly affect the above-threshold behavior of laser systems. They are 
induced by pumping the laser nonuniformly \cite{ds}. In the vicinity of the EPs, where the
laser modes coalesce, the effect of the singularities can explain the turning off of one laser
even when the overall pump power deposited in the system is increased. 

The relation between EPs in closed and open gain-loss structures
has been expounded recently \cite{SR}. The EPs occurring for a closed-system Hamiltonian are 
related to those of the scattering matrix for the corresponding open system. It has been demonstrated in
detail how these seemingly different situations share a very close and elegant connection with each
other.

\begin{figure}
\includegraphics[height=0.30\textheight]{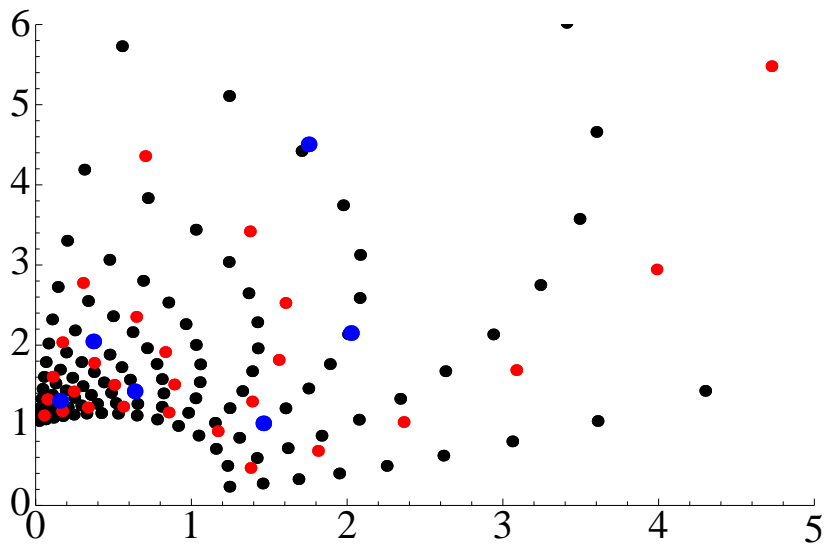}
\caption{Exceptional points in the complex $\la $-plane for the Lipkin model
 with $N=8$ (blue), $N=16$ (red) and $N=32$ (black). The symmetry of the model
yields EPs in the other quadrants: with $\la _{EP}$ additional EPs occur
at  $\la _{EP}^{*}$ and $-\la _{EP}$. }
\end{figure}

\subsection{Quantum phase transitions, chaos}
The important role played by EPs in connection with quantum phase transition 
and also quantum chaos in many body systems 
appears to be lesser appreciated by the 'EP-community'. We here report
some pertinent results. 
The Lipkin model \cite{lip} is a toy model often used to study quantum phase transitions
of many body systems.
The interaction of the two level model lifts or lowers a Fermion pair between the
two levels. For $N$ particles it can be formulated in terms of the angular 
momentum operators and reads
\bea
H(\la )=J_z+\frac{\la }{N}(J_+^2+J_-^2)
\eea
with $J_z,J_{\pm}$ being the $N$-dimensional representations of the $SU(2)$
operators. There is a phase transition at $\la >1$ that moves toward $\la =1$ in
the thermodynamic limit ($N\to \infty$). The special interaction gives rise to an 
inherent symmetry: even and odd numbers $k$ of the ordered levels $E_k$ do not interact. 
The phase for $\la <1$ is the 'normal' phase where
the symmetry of the problem is preserved by the levels and wave functions.
In the 'deformed' phase for $\la >1$ the symmetry is broken in that
even and odd $k$ become degenerate. Here the role of the EPs is
crucial to bring about the phase transition in the spectrum \cite{gehe,ley}. In Fig.2 the
pattern of the EPs is illustrated for low values of $N$. It is clearly
seen how the EPs accumulate for increasing $N$ on the real axis with
the tendency to move towards the point $\la =1$. The spectrum remains
unaffected by singularities in the region of the normal phase while
it is strongly affected around the critical point. For finite
temperature these singularities feature in the partition
function as is discussed explicitly in \cite{cej}. If the model is perturbed the regular pattern of
the EPs is destroyed and so is the spectrum accordingly. The onset of
chaos \cite{hesa} is clearly discernible in the region of the phase transition
associated with a high density of EPs
while the model remains robust outside the critical region for
sufficiently mild perturbation.

\subsection{Special effects in multichannel scattering}
Depending on a judicious choice of parameters the proximity of EPs
can invoke dramatic effects in multichannel scattering such as a sudden
increase of the cross section in one channel, even by orders of magnitude.
In turn, a second channel is suppressed and can show a resonance curve that
deviates substantially from the usual Lorentz shape \cite{HN10}. Related
to this behaviour is the pattern in the time domain \cite{H10}. Depending
on the initial conditions the wave function displays characteristic features
such as very fast decay or the opposite, i.e.~very long life time. 
At the EP the time dependent wave function typically has a linear term in
time besides the usual exponential behaviour.

\subsection{EPs of higher order}
The coalescence of three levels \cite{sokol} (or more) is of course possible if sufficiently
many parameters are at one's disposal. For an EPN ($N$ levels coalescing) an
$N-$dimensional matrix must be considered. For complex symmetric matrices
$(N^2+N-2)/2$ parameters are needed to enforce the coalescence of $N$ levels \cite{he3}.
An interesting aspect is the behaviour of, say, an EP3 where two EP2 sprout out from the EP3
under variation of one of the parameters; in this way we can view an EP3 as the
coalescence of two EP2 being linked by their position on one common Riemann
sheet. This can be generalised allowing many combinations for larger $N$: the
block structure of the Jordan form of the Hamiltonian gives an indication
of the connectedness of the $N$ levels.
In a recent paper \cite{evg} this is investigated for EP3;
a physical example for possible implementation in the laboratory is suggested.
Topological properties and their group structure of higher order
EPs are investigated in \cite{kim}.

\subsection{Classical systems}
Even though some of the experiments with microwave cavities as well as
some of the optical systems are classical in nature we here note in particular
effects of EPs in classical \cite{mely} and fluid mechanics. Instabilities and
a particular behaviour of the Reynold number for a Poiseuille flow have been associated
with near crossings of eigenvalues and -- as is now identified  -- with EPs
\cite{cajones,or} (see also \cite{buss}).

\section{The role of EPs in approximation schemes}
The well known Random Phase Approximation (RPA) used in many body problems
yields an effective Hamiltonian that is
non-hermitian \cite{ring}. As a result, eigenvalues are not necessarily real. Depending on the
strength of the, say, particle-hole interaction two real eigenvalues $\Omega$ and
$-\Omega$ coalesce at $\Omega=0$ and then move into the complex plane when the interaction
is increased. Often this instability point is associated with one or more phase transition
of the underlying mean field \cite{hena}. It is an EP with all its characteristics:
square root branch point in the interaction strength and the vanishing norm of the
wave function.

A perturbative approach in shell model calculations can be hampered by singularities
associated with {\em intruder states} \cite{wesu}. These singularities are EPs
where two levels coalesce thus limiting the radius of convergence of the perturbation
series. In a similar vein, branch point singularities (EPs) have been identified
to affect the convergence of perturbation series in a field theoretical
model \cite{CMB1} and for the anharmonic oscillator \cite{CMB2}; in these infinite dimensional
cases accumulation points of EPs are encountered.

Recent approaches to model nuclei near to the drip line \cite{naz} use resonance states
to describe the continuum. The coalescence of two resonances can invoke specific
physical effects owing to the strong increase of the associated spectroscopic
factors being caused by the vanishing norm of the wave functions at the EP.

\section{Summary}
The ubiquitous occurrence of EPs in all eigenvalue problems that depend
on a parameter can have significant and often dramatic effects on
observables in a great variety of physical phenomena. A few decades ago,
these singularities appeared as a purely mathematical feature that could cause
problems in approximation schemes. It was only about fifteen years ago that their
physical manifestation has been demonstrated in experiments that were basically
classical in nature. At present definite theoretical and
experimental proposals are found in the literature relating to
atomic and molecular physics, using lasers for
triggering and measuring specific transitions. In nuclear physics, where
there is now great interest in open systems, that is in nuclei on the drip line,
the coalescence of resonance states is expected to produce specific effects
such as enhancements of particular reactions. The tremendous up-beat of research
in connection with ${\cal PT}$-symmetric Hamiltonians is expected to further proliferate
new effects and applications of EPs also and in particular in genuine quantum
mechanical problems.

\end{document}